\documentclass[aps,prb,twocolumn,superscriptaddress]{revtex4-2}

\usepackage[usenames,dvipsnames]{color}

\usepackage{comment}
\usepackage{graphicx}
\usepackage[utf8x]{inputenc}
\usepackage{lineno}
\usepackage{color}
\usepackage{hyperref}
\usepackage{amsmath}
\usepackage{bm}
\usepackage{braket}
\usepackage{caption}
\usepackage{subcaption}

\newcommand\cdks[2]{c^{\dagger}_{k#1,#2\sigma}}

\newcommand\cks[2]{c_{k#1,#2\sigma}}

\begin{document}

\title{Planckian behaviour in the optical conductivity of the weakly coupled Hubbard model}
\author{Maxence Grandadam}
\affiliation{Department of Physics and Physical Oceanography, Memorial University of Newfoundland, St. John's, Newfoundland \& Labrador, Canada A1B 3X7} 
\author{J. P. F. LeBlanc}
\email{jleblanc@mun.ca}
\affiliation{Department of Physics and Physical Oceanography, Memorial University of Newfoundland, St. John's, Newfoundland \& Labrador, Canada A1B 3X7}

\date{\today}
\begin{abstract}
We study the frequency and temperature dependence of the optical conductivity in the weakly coupled two-dimensional Hubbard model using a renormalized perturbative expansion. The perturbative expansion is based on the skeleton series for the current-current correlation function with a dressed Green`s function and the results are obtained directly on the real frequency axis using Algorithmic Matsubara Integration (AMI). The resulting conductivity shows a temperature-independent power law behaviour in the intermediate frequency regime. Moreover, the associated transport scattering time and renormalized mass exhibit a Planckian behaviour. We show that the self-energy of the Hubbard model, however, is distinct from existing Planckian models. The Planckian behaviour of the conductivity, observed in optimally doped cuprates for example, can thus be obtained from a different form of self-energy than the Planckian model, such as the weakly coupled Hubbard model at half-filling.
\end{abstract}

\maketitle

\section{Introduction}
Strange metals are a key problem of interest in the study of strongly correlated systems for the past years. The wide variety of experimental observations of strange metallicity\cite{gurvitch1987resistivity,SMTBG20,StewartRMP,rost11_Ruth} and the number of associated theoretical descriptions\cite{VarmaMFL,chubukov_adv_phys,tsvelik2017ladder,rice2017umklapp,hartnoll2018holographic,wu2018candidate,patel2018magnetotransport,ErezBerg20,Paul13,McKenzieDMFT00,huang19strange} makes it difficult to extract a unique underlying principle. The main characteristic of the strange metal regime is the linear temperature dependence of the longitudinal resistivity in contrast to the quadratic dependence expected from Fermi liquid theory. More recently, a study of the optical conductivity in a cuprate material has also shown signatures of this strange metal behaviour in the frequency dependence of the longitudinal conductivity\cite{Michon}. The phenomenology was well reproduced by assuming a form for the self-energy of the system associated with a ``Planckian model'' usually obtained from SYK or Kondo-like models.

We present here a study of the optical conductivity in the weakly coupled 2D Hubbard model from a normalized perturbative expansion of the current-current correlation function. We use Algorithmic Matsubara Integration\cite{AMI,libami} (AMI) to compute the current-current correlation function directly on the real frequency axis without the need for numerical analytic continuation and extract from it the transport scattering time and mass renormalization by using a generalized Drude model. The results for the conductivity present the same behaviour in frequency and temperature as both the experimental and theoretical works even though the self-energy of the Hubbard model does not satisfy the constraint given in the Planckian model. We emphasize the importance of the momentum dependence of the self-energy to the final result and discuss the role of vertex corrections.

\section{Scaling of the optical conductivity}
We study the half-filled, weakly coupled, Hubbard model
\begin{eqnarray}\label{E:Hubbard}
H = \sum_{ ij \sigma} t_{ij}c_{i\sigma}^\dagger c_{j\sigma} + U\sum_{i} n_{i\uparrow} n_{i\downarrow},
\end{eqnarray}
where $t_{ij}$ is the hopping amplitude, $c_{i\sigma}^{(\dagger)}$ ($c_{i\sigma}$) is the creation (annihilation) operator at site $i$, $\sigma \in \{\uparrow,\downarrow\}$ is the spin, $U$ is the onsite Hubbard interaction, $n_{i\sigma} = c_{i\sigma}^{\dagger}c_{i\sigma}$ is the number operator.  We restrict the sum over sites to nearest neighbours for a 2D square lattice at half-filling and use $U=3$ unless stated otherwise in the following.
We use a skeleton series expansion for the current-current correlation function $\chi_{jj}(i\omega_n) = -\frac{1}{V}\int d\tau e^{i\omega_n\tau}\left\langle T_{\tau}j\left(\tau\right)j\left(0\right)\right\rangle$, where $j = \sum_{k,\sigma} v_k \cdks{}{}\cks{}{}$ is the current operator, from which we obtain the optical conductivity $\sigma(\omega) = i\left(\chi_{jj}(0)-\chi_{jj}(\omega)\right)/\omega$ (see supplemental material for details ). Using AMI we can perform the analytic continuation symbolically without resorting to numerical techniques such as maximum entropy inversion MAXENT\cite{jarrell:maxent,maxent}, Pade approximants\cite{pade} or other modern methods\cite{gull:nevanlinna,Nogaki2023}. We thus obtain the optical conductivity at different temperatures between $\beta=20$ and $\beta=2$.\\

\begin{figure}
	\centering
	\includegraphics[width=\linewidth]{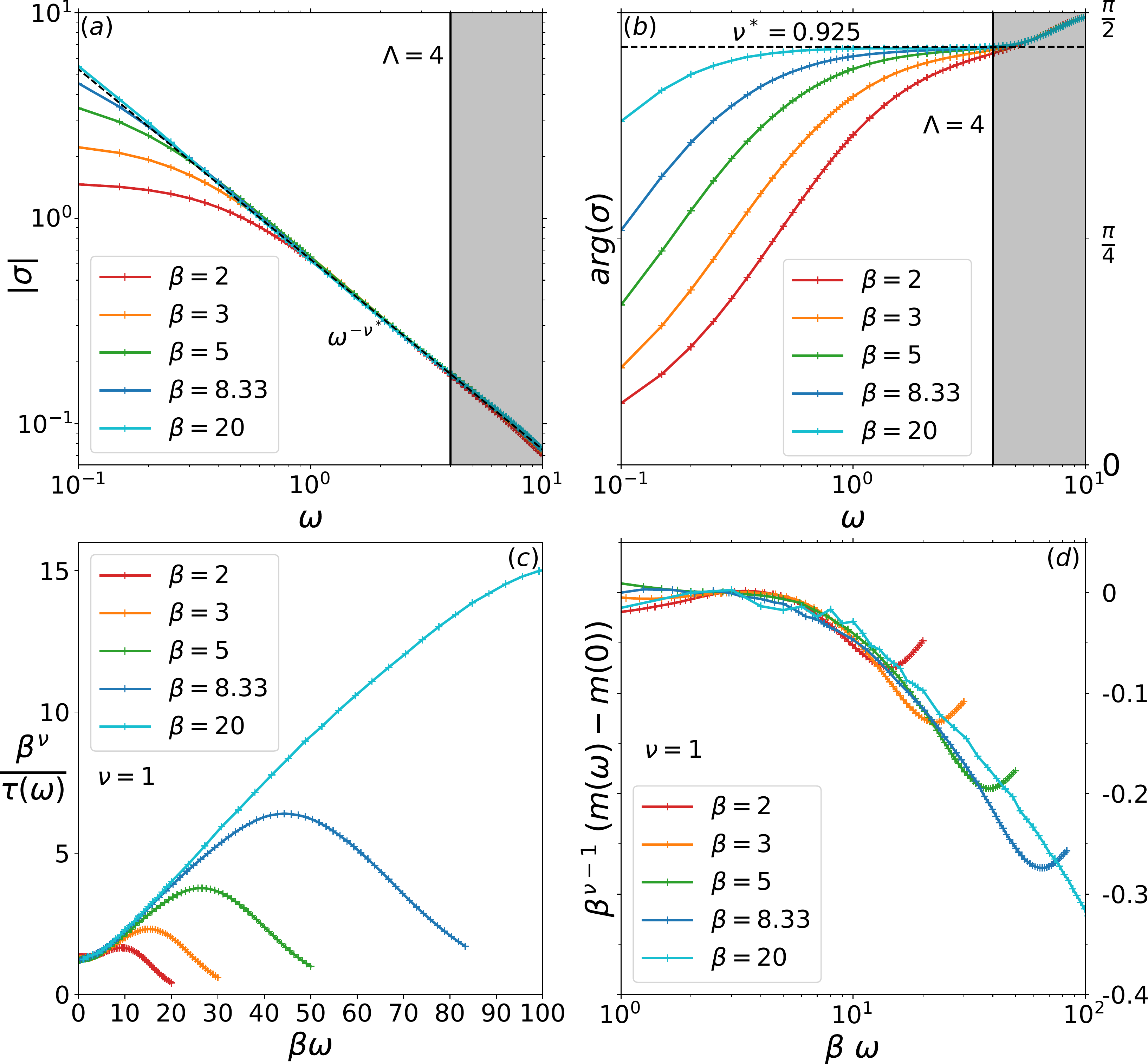}
	\caption{Frequency dependence of the modulus $\bm{(a)}$ and phase $\bm{(b)}$ of the conductivity at different temperature. The dashed line indicates the power law form of the modulus $\omega^{\nu^*}$ and the saturation bound $\nu^* \frac{\pi}{2}$ for the phase respectively. The exponent $\nu^*$ is temperature independent. The high energy cut-off $\Lambda$ is taken as the frequency at which the phase deviates from this bound. $\bm{(c)}$ inverse scattering time and $\bm{(d)}$ effective mass extracted from the conductivity when rewritten as an extended Drude model given in Eq.\eqref{eq:drude2}. The frequency axis is scaled by the inverse temperature for each curve as well as the scattering time to exhibit the linear scaling obeyed by the two quantities. An approximate $m(0)$ has been subtracted from the effective mass but is subject to the important noise in the data.}
	\label{fig:fig2}
\end{figure}

The modulus and phase of the optical conductivity are presented in Fig.\ref{fig:fig2}(a) and (b) and we can clearly identify three different frequency regimes. At low frequencies, the modulus goes to a temperature-dependent constant while for frequencies in the intermediate regime $T \ll \omega \ll \Lambda$, the modulus of the conductivity exhibits a power law dependence $|\sigma\left(\omega\right)| \propto \omega^{-\nu^*}$ with a temperature-independent exponent $\nu^* < 1$. In the same frequency range, the phase of the conductivity saturates at $\nu^* \pi/2$ lower than the expected $\pi/2$ which is only reached at higher frequencies above a high-energy cut-off $\Lambda$. It is important to note that the dc conductivity $\sigma\left(\omega \to 0\right)$ exhibits a $1/T$ temperature dependence consistent with the linear-in-$T$ behaviour observed in other work on the Hubbard model at strong\cite{Huang2019} and at weak\cite{Vucicevic2022} coupling (see supplemental material).\\

To understand the frequency dependence of the conductivity, it is useful to rewrite it in a generalized Drude form with a frequency-dependent scattering rate and optical mass
\begin{equation}
\sigma\left(\omega\right) = \frac{\sigma_0}{1/\tau\left(\omega\right)-i \omega \frac{m^*\left(\omega\right)}{m_0}}\label{eq:drude2}.
\end{equation}
These two quantities can then be extracted from the conductivity data by inverting Eq.\eqref{eq:drude2} leading to
\begin{equation}
\frac{1}{\tau\left(\omega\right)} = \text{Re}\left[\frac{1}{\sigma\left(\omega\right)}\right], \qquad
\frac{m^*\left(\omega\right)}{m_0} = -\text{Im}\left[\frac{1}{\omega \sigma\left(\omega\right)}\right] \label{eq:mass}.
\end{equation}
Extracting the scattering time and the effective mass from the conductivity and rescaling the former by the inverse temperature reveals the temperature scaling of the two quantities as shown in Fig.\ref{fig:fig2}(c) and (d). The scaling of the scattering time at low frequency with the inverse temperature is representative of the linear dependence of the dc conductivity. The collapse of curves at intermediate frequencies in Fig.\ref{fig:fig2}(c) and (d) is the hallmark of Planckian behaviour.

It was pointed out recently\cite{Michon} that both the power law behaviour of the conductivity with $\nu^* < 1$ and the scaling of the scattering time with $\nu=1$ we obtained in the Hubbard model can be obtained from a single ``Planckian model'' where one forces the conductivity to have Planckian behaviour by a particular choice of self-energy. In that Planckian model, there is a direct relation between the value of the interaction and the exponent $\nu^*$ and one can show that this is also satisfied in the Hubbard model (see supplemental material). We discussed this Planckian model ansatz for the self-energy that leads to a Planckian behaviour of the conductivity and compare it with the self-energy obtained in our calculation of the 2D Hubbard model.

\section{Planckian model and local self-energy}
\subsection{Basics of the Planckian ansatz}
One way of obtaining a linear temperature dependence for the resistivity is to assume a form of the self-energy that has the right temperature dependence as $\omega \to 0$. In order to give the observed scaling at intermediate frequencies in the scattering time, the self-energy should also be a function of the ratio $\omega/T$. These self-energy ansatzes are usually referred to as (sub-)Planckian models and are given by\cite{Michon}
\begin{align}
&\text{Im}\left[ \Sigma \left( \omega \right) \right] = -\frac{g\pi}{\beta^{\nu}} \ f \left( \beta \omega \right) \label{eq:planckian},
\end{align}
where $f\left(x\right)$ is a function that goes to a constant at low frequency, $f\left(x \to 0 \right) \to f_0$, and has a power law form at high frequency, $f\left(x \gg 1 \right) \to |x|^{\nu}$. Note that this form of self-energy is local, \emph{i.e.} does not have any momentum dependence, which is usually justified by the weak momentum dependence observed experimentally in cuprates\cite{Millis2003,Grissonnanche2021}.

The self-energy in Eq.\eqref{eq:planckian} with $\nu=1$ naturally leads to a linear temperature dependence of the dc conductivity as well as the scaling observed in the scattering time and in the effective mass. The ansatz for the local Planckian self-energy thus seems to give a satisfactory description of the results obtained both experimentally and from our study of the Hubbard model. One of the advantages of the Planckian model is that the local nature of the self-energy leads to a cancellation of the vertex correction in the Kubo formula for the conductivity\cite{Khurana1990,Vucicevic2021} which would make the low-order expansion used in this work sufficient for a full description of the conductivity. It is known that such a form of the self-energy can be obtained from microscopic models such as SYK-type models\cite{Dumitrescu2021,Patel2022} or Kondo-like models\cite{Parcollet1998}. These models usually rely on some large-N approach to obtain the linear scaling of the self-energy in a non-perturbative scheme. The fact that the Hubbard model can exhibit such properties at second-order in perturbation theory may thus seem surprising.

\subsection{Breakdown of the local assumption}
\begin{figure}
	\centering
	\includegraphics[width=0.9 \linewidth]{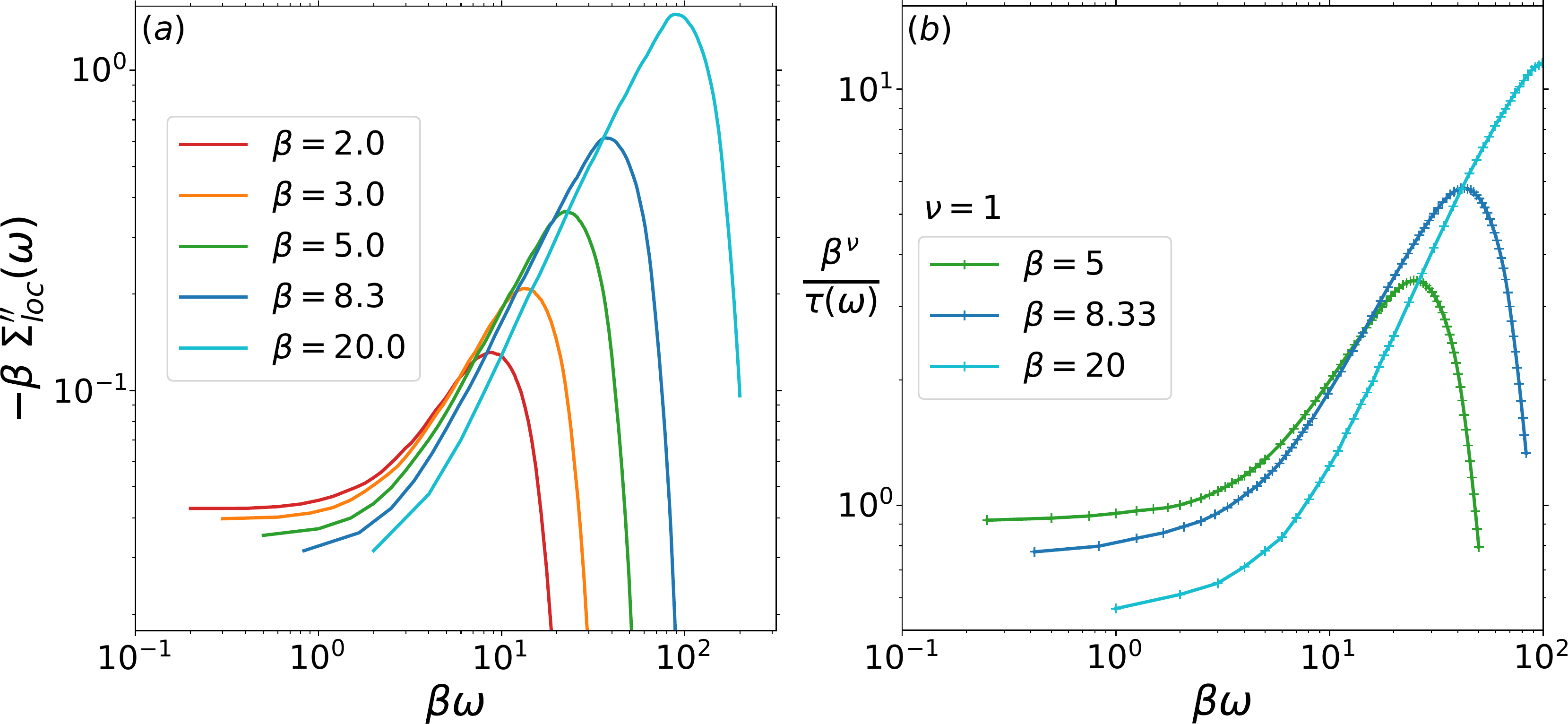}
	\caption{$\bm{(a)}$ The Scaling of the imaginary part of the local self-energy $\Sigma_{\text{loc}}$ at different temperatures. The absence of collapse of the data shows that the local self-energy does not obey the same scaling as the Planckian ansatz Eq.\eqref{eq:planckian}. $\bm{(b)}$ The Scaling of the scattering time obtained using the local self-energy shown in panel $(a)$. In contrast to Fig.\ref{fig:fig2}, there is no scaling of the scattering time with temperature.}
	\label{fig:fig4}
\end{figure}

Each calculation of the conductivity required us to compute the self-energy on a grid in momentum and frequency space from which we can directly check and compare the Planckian ansatz presented in Eq.\eqref{eq:planckian}. One popular assumption is to consider a local self-energy obtained from integrating the momentum-dependent results at second-order in perturbation
\begin{equation}
\Sigma_{\text{loc}}\left(\omega\right) = \frac{1}{N}\sum_{k}\Sigma\left(k,\omega\right).
\end{equation}
The local approximation, as is the case in methods such as dynamical mean-field theory (DMFT), has been shown to lead to a linear-in-temperature resistivity over a wide range of temperatures\cite{Cha2020}. The resulting imaginary part for the local self-energy in our calculation is shown in Fig.\ref{fig:fig4}(a) for positive frequencies. The large energy cut-off $\Lambda$ identified previously in the conductivity is related to the bandwidth after which the amplitude of the self-energy decreases strongly.

Scaling the imaginary part of the self-energy as indicated in Eq.\eqref{eq:planckian} does not produce any collapse of the calculations at each temperature as shown in Fig.\ref{fig:fig4}(a). In fact, there is no value of $\nu$ that leads to a collapse of the data (see supplemental material). This is an indication that the local form of the self-energy obtained by integrating the momentum-dependent second-order self-energy for the Hubbard model does not fall into the category of Planckian models.

It is important to remember that the local self-energy presented in Fig.\ref{fig:fig4}(a) is not the one that was used to obtain the conductivity in Fig.\ref{fig:fig2} which included the full momentum dependence. In fact, the scattering time extracted from the calculation using the local form of the self-energy is presented in Fig.\ref{fig:fig4}(b) and does not show the same scaling behaviour as the original results. The resolution to this apparent paradox comes from the momentum dependence of the self-energy in the Hubbard model with a peculiar differentiation between the nodal and the antinodal point.

\section{Nodal - Antinodal dichotomy}
\begin{figure}
	\centering
	\includegraphics[width=\linewidth]{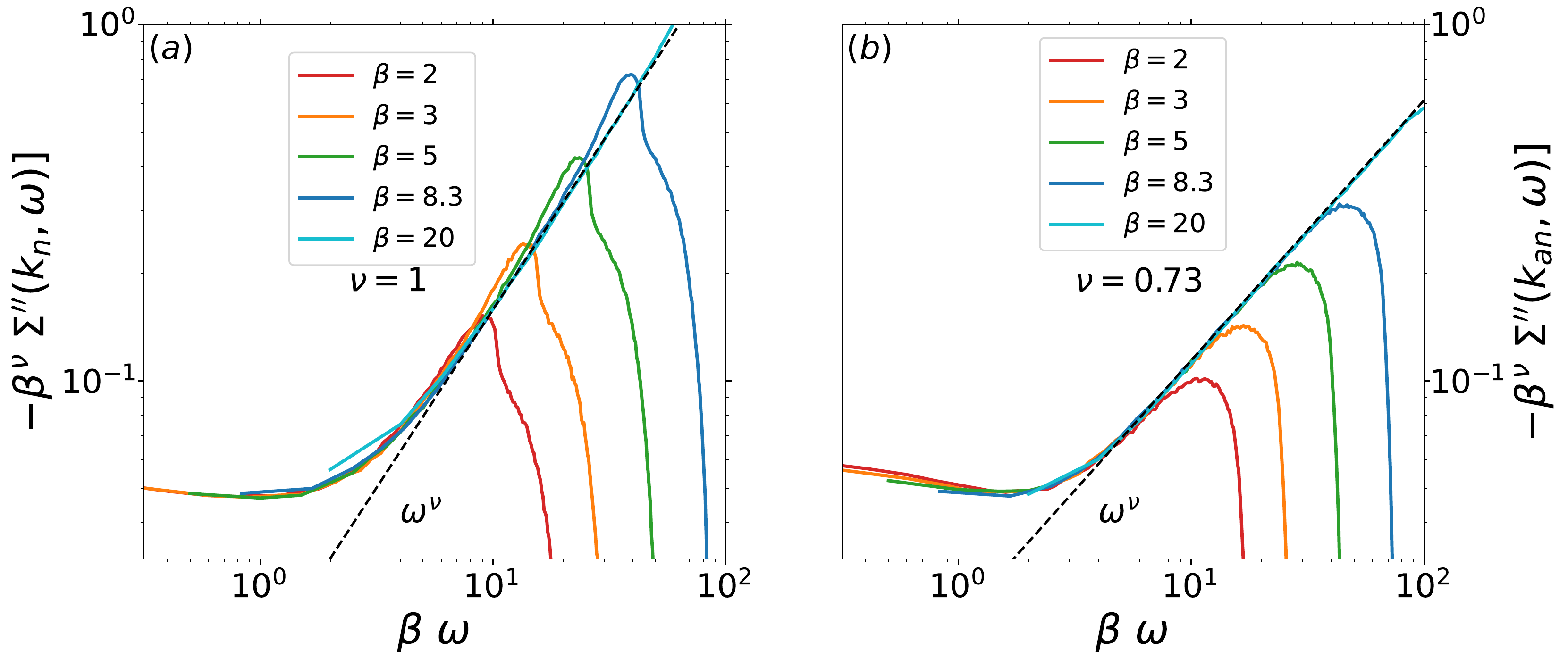}
	\caption{$\bm{(a)}$ Scaling of the imaginary part of the self-energy at the nodal point $k_n = \left(\frac{\pi}{2},\frac{\pi}{2}\right)$ that shows a Planckian behaviour with $\nu=1$. $\bm{(b)}$ Scaling of the imaginary part of the self-energy at the antinodal point $k_{an} = \left(\pi,0\right)$ that shows a sub-Planckian behaviour with $\nu<1$.}
	\label{fig:fig5}
\end{figure}

It was pointed out in previous work that the second-order self-energy of the half-filled Hubbard model takes a ``marginal Fermi liquid''(MFL) form at low temperature\cite{Kakehashi2005,Virosztek1990,Czycholl1991}. The main reason for this seems to be the presence of the van Hove singularity in the density of states that leads to an anomalous frequency dependence of the self-energy on the Fermi surface. The typical form for the imaginary part of the self-energy in the MFL case is given by $\text{Im}\left[\Sigma_{\text{MFL}}\left(\omega\right)\right] = -g \ \pi \ \text{max}\left(|\omega|,T\right)$
which has the same scaling properties as the self-energy from the Planckian model Eq.\eqref{eq:planckian} as $\Sigma_{\text{MFL}}\left(\omega \to 0 \right) \sim T$ and  $\Sigma_{\text{MFL}}\left(\omega \gg T \right) \sim |\omega|$. The self-energy we computed does in fact reproduce this behaviour at the nodal point $k_{n}=\left(\frac{\pi}{2},\frac{\pi}{2}\right)$ as shown in Fig.\ref{fig:fig5}(a). Once again we see that the imaginary part of the self-energy exhibit a linear power law dependence in the intermediate frequency regime $T \ll \omega \ll \Lambda$ and goes to a constant at low frequencies that scales as $\Sigma\left(k_n,\omega \to 0 \right) \propto T$.

Trying to find a similar scaling for the self-energy in the antinodal region, $k_{an} = \left(\pi,0\right)$, however, leads to a different result as shown in Fig.\ref{fig:fig5}(b). The self-energy now takes the form of a ``sub-planckian'' model from Eq.\eqref{eq:planckian} with an exponent $\nu<1$. This is the reason for the apparent breakdown of the Planckian behaviour of the local self-energy as we showed in Fig.\ref{fig:fig4}.

In the case of the half-filled Hubbard model and without vertex corrections, the band velocity $\partial \epsilon_k/ \partial k_x$ that appears in the conductivity formula is larger at the nodal point and vanishes at the antinode. This could be an explanation for the apparent Planckian behaviour observed in the conductivity despite the sub-Planckian behaviour observed at the antinode.

\section{Effect of vertex corrections}
\begin{figure}
	\centering
	\includegraphics[width=0.9 \linewidth]{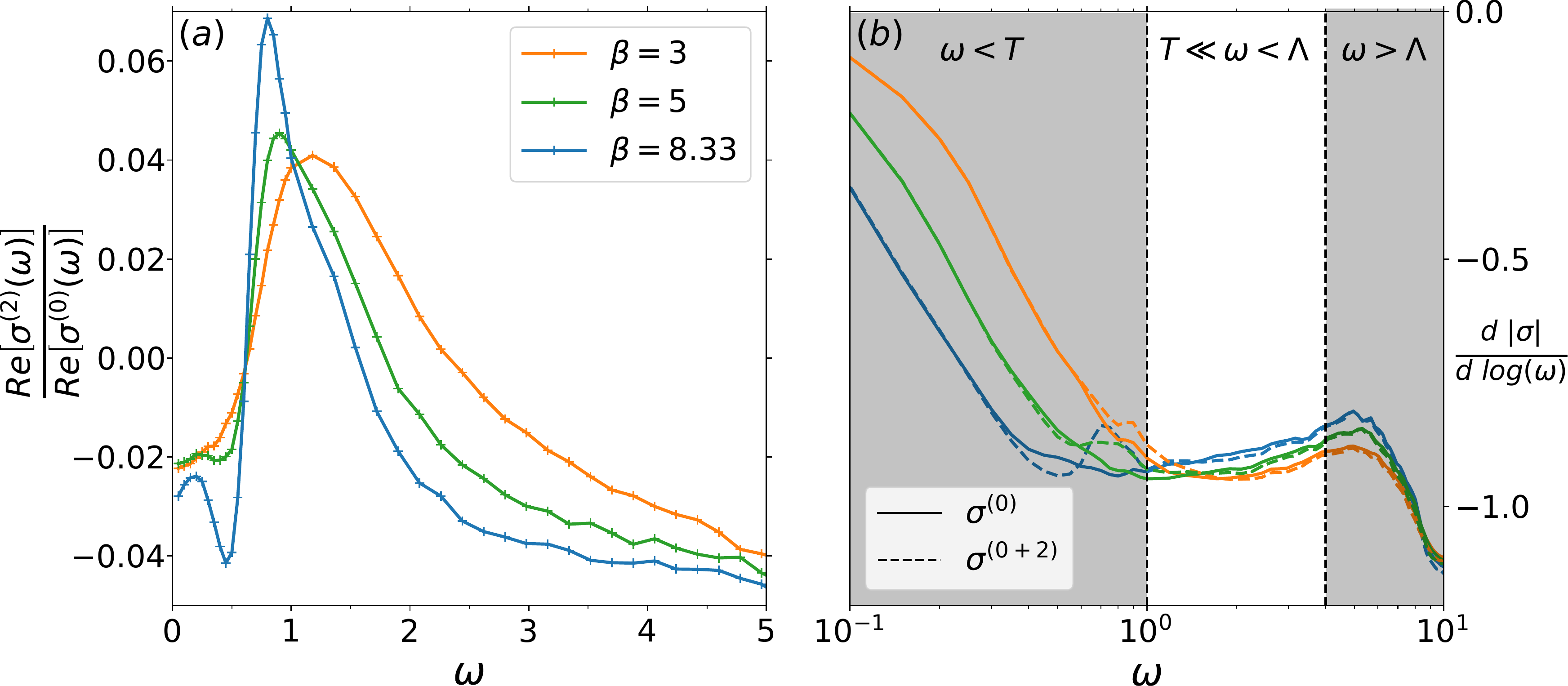}
	\caption{$\bm{(a)}$ Relative contribution of the vertex corrections with respect to the zeroth order contribution to the real part of the conductivity at different temperatures. At these temperatures, the vertex corrections are small and mainly affect the low-frequency regime. $\bm{(b)}$ Derivative of the modulus of the conductivity without (full lines) and with (dashed lines) the inclusion of vertex corrections. The main changes are shown to occur at low frequencies and do not affect the power law scaling discussed previously in the intermediate frequency regime $T \ll \omega < \Lambda$.}
	\label{fig:fig6}
\end{figure}

In light of the strong momentum dependence of the scaling properties of the self-energy presented in the previous section, it is important to take into account the vertex corrections to the conductivity. These second-order contributions to the conductivity are shown in Fig.\ref{fig:fig6}(a) for different temperatures. The overall amplitude for the vertex correction is much smaller than the one obtained at the leading order (bubble diagram). 

In fact, the second-order term accounts for less than $10\%$ of the zeroth-order contribution to the total conductivity for the three temperatures presented here. The importance of the vertex correction grows when lowering the temperature and as such, they may become significant at lower temperatures. Nonetheless, the main impact of the vertex correction is seen at low frequency, outside of the intermediate frequency regime where the scaling properties of the conductivity we discussed previously occur. This is apparent in Fig.\ref{fig:fig6}(b) where the logarithmic derivative of the modulus of the conductivity is shown for different temperatures. We thus expect our previous conclusion to hold even when vertex corrections, related to the momentum dependence of the self-energy, are included.

\section{Conclusion}

We studied the frequency dependence of the longitudinal conductivity using a renormalized Green functions scheme. This allows us to obtain results directly on the real-frequency axis without any numerical analytical continuation and without the need for finite $\Gamma$ in the $i\omega_n \to \omega + i\Gamma$ analytic continuation of AMI integrands\\
\\
At leading order, the conductivity for the weakly coupled Hubbard model shows a power law behaviour at intermediate frequencies $T \ll \omega \ll \Lambda$ with a temperature independent exponent $\nu^{*}<1$. This is reminiscent of the behaviour of the optical conductivity in strange metals such as cuprates. Moreover, it was recently shown that Planckian models can lead to such power law scaling\cite{Michon}. The frequency dependence of the scattering time and effective mass obtained by fitting the conductivity to an extended Drude form do, in fact, show Planckian behaviour with notably $1/\tau\left(\omega\right) \propto \beta \omega$.\\
\\
The self-energy in the Hubbard model is however significantly different from the local self-energy ansatz of the Planckian models. Even when getting rid of the momentum dependence by averaging the self-energy over all momenta, the resulting local quantity does not exhibit Planckian behaviour, and the resulting conductivity also has a different temperature dependence. The momentum dependence of the self-energy in the Hubbard model is thus crucial to obtain the power law scaling observed in the scattering time and effective mass. We showed that the vertex corrections that arise from this momentum dependence do not affect our conclusion as they mostly affect the low-frequency part of the conductivity.\\
\\
The results presented here are based on a perturbative expansion and are thus expected to be relevant in the weak coupling regime. It is surprising that features such as Planckian behaviour of the conductivity, usually associated with non-Fermi liquids and strongly correlated systems, can be recovered for small $U$. The half-filled Hubbard model in itself is far from the complexity that exists in real cuprates materials and as such, it would be interesting to explore the doped case with next-nearest neighbours hoping. Previous studies using Cellular Dynamical Mean Field Theory have in fact shown that an extended region of linear-in-temperature resistivity\cite{Wei2022} can be observed in the doped Hubbard model.\\

We acknowledge the support of the Natural Sciences and Engineering Research Council of Canada (NSERC) RGPIN-2022-03882 and support from the Simons Collaboration on the Many Electron Problem.

\bibliographystyle{apsrev4-1}

\bibliography{library18_08.bib}

\begin{thebibliography}{37}%
\makeatletter
\providecommand \@ifxundefined [1]{%
 \@ifx{#1\undefined}
}%
\providecommand \@ifnum [1]{%
 \ifnum #1\expandafter \@firstoftwo
 \else \expandafter \@secondoftwo
 \fi
}%
\providecommand \@ifx [1]{%
 \ifx #1\expandafter \@firstoftwo
 \else \expandafter \@secondoftwo
 \fi
}%
\providecommand \natexlab [1]{#1}%
\providecommand \enquote  [1]{``#1''}%
\providecommand \bibnamefont  [1]{#1}%
\providecommand \bibfnamefont [1]{#1}%
\providecommand \citenamefont [1]{#1}%
\providecommand \href@noop [0]{\@secondoftwo}%
\providecommand \href [0]{\begingroup \@sanitize@url \@href}%
\providecommand \@href[1]{\@@startlink{#1}\@@href}%
\providecommand \@@href[1]{\endgroup#1\@@endlink}%
\providecommand \@sanitize@url [0]{\catcode `\\12\catcode `\$12\catcode
  `\&12\catcode `\#12\catcode `\^12\catcode `\_12\catcode `\%12\relax}%
\providecommand \@@startlink[1]{}%
\providecommand \@@endlink[0]{}%
\providecommand \url  [0]{\begingroup\@sanitize@url \@url }%
\providecommand \@url [1]{\endgroup\@href {#1}{\urlprefix }}%
\providecommand \urlprefix  [0]{URL }%
\providecommand \Eprint [0]{\href }%
\providecommand \doibase [0]{http://dx.doi.org/}%
\providecommand \selectlanguage [0]{\@gobble}%
\providecommand \bibinfo  [0]{\@secondoftwo}%
\providecommand \bibfield  [0]{\@secondoftwo}%
\providecommand \translation [1]{[#1]}%
\providecommand \BibitemOpen [0]{}%
\providecommand \bibitemStop [0]{}%
\providecommand \bibitemNoStop [0]{.\EOS\space}%
\providecommand \EOS [0]{\spacefactor3000\relax}%
\providecommand \BibitemShut  [1]{\csname bibitem#1\endcsname}%
\let\auto@bib@innerbib\@empty
%</preamble>
\bibitem [{\citenamefont {Gurvitch}\ and\ \citenamefont
  {Fiory}(1987)}]{gurvitch1987resistivity}%
  \BibitemOpen
  \bibfield  {author} {\bibinfo {author} {\bibfnamefont {M.}~\bibnamefont
  {Gurvitch}}\ and\ \bibinfo {author} {\bibfnamefont {A.~T.}\ \bibnamefont
  {Fiory}},\ }\href {\doibase 10.1103/PhysRevLett.59.1337} {\bibfield
  {journal} {\bibinfo  {journal} {Phys. Rev. Lett.}\ }\textbf {\bibinfo
  {volume} {59}},\ \bibinfo {pages} {1337} (\bibinfo {year}
  {1987})}\BibitemShut {NoStop}%
\bibitem [{\citenamefont {Cao}\ \emph {et~al.}(2020)\citenamefont {Cao},
  \citenamefont {Chowdhury}, \citenamefont {Rodan-Legrain}, \citenamefont
  {Rubies-Bigorda}, \citenamefont {Watanabe}, \citenamefont {Taniguchi},
  \citenamefont {Senthil},\ and\ \citenamefont {Jarillo-Herrero}}]{SMTBG20}%
  \BibitemOpen
  \bibfield  {author} {\bibinfo {author} {\bibfnamefont {Y.}~\bibnamefont
  {Cao}}, \bibinfo {author} {\bibfnamefont {D.}~\bibnamefont {Chowdhury}},
  \bibinfo {author} {\bibfnamefont {D.}~\bibnamefont {Rodan-Legrain}}, \bibinfo
  {author} {\bibfnamefont {O.}~\bibnamefont {Rubies-Bigorda}}, \bibinfo
  {author} {\bibfnamefont {K.}~\bibnamefont {Watanabe}}, \bibinfo {author}
  {\bibfnamefont {T.}~\bibnamefont {Taniguchi}}, \bibinfo {author}
  {\bibfnamefont {T.}~\bibnamefont {Senthil}}, \ and\ \bibinfo {author}
  {\bibfnamefont {P.}~\bibnamefont {Jarillo-Herrero}},\ }\href@noop {}
  {\bibfield  {journal} {\bibinfo  {journal} {Physical Review Letters}\
  }\textbf {\bibinfo {volume} {124}},\ \bibinfo {pages} {076801} (\bibinfo
  {year} {2020})}\BibitemShut {NoStop}%
\bibitem [{\citenamefont {Stewart}(2001)}]{StewartRMP}%
  \BibitemOpen
  \bibfield  {author} {\bibinfo {author} {\bibfnamefont {G.~R.}\ \bibnamefont
  {Stewart}},\ }\href {\doibase 10.1103/RevModPhys.73.797} {\bibfield
  {journal} {\bibinfo  {journal} {Rev. Mod. Phys.}\ }\textbf {\bibinfo {volume}
  {73}},\ \bibinfo {pages} {797} (\bibinfo {year} {2001})}\BibitemShut
  {NoStop}%
\bibitem [{\citenamefont {Rost}\ \emph {et~al.}(2011)\citenamefont {Rost},
  \citenamefont {Grigera}, \citenamefont {Bruin}, \citenamefont {Perry},
  \citenamefont {Tian}, \citenamefont {Raghu}, \citenamefont {Kivelson},\ and\
  \citenamefont {Mackenzie}}]{rost11_Ruth}%
  \BibitemOpen
  \bibfield  {author} {\bibinfo {author} {\bibfnamefont {A.}~\bibnamefont
  {Rost}}, \bibinfo {author} {\bibfnamefont {S.~A.}\ \bibnamefont {Grigera}},
  \bibinfo {author} {\bibfnamefont {J.}~\bibnamefont {Bruin}}, \bibinfo
  {author} {\bibfnamefont {R.}~\bibnamefont {Perry}}, \bibinfo {author}
  {\bibfnamefont {D.}~\bibnamefont {Tian}}, \bibinfo {author} {\bibfnamefont
  {S.}~\bibnamefont {Raghu}}, \bibinfo {author} {\bibfnamefont {S.~A.}\
  \bibnamefont {Kivelson}}, \ and\ \bibinfo {author} {\bibfnamefont
  {A.}~\bibnamefont {Mackenzie}},\ }\href@noop {} {\bibfield  {journal}
  {\bibinfo  {journal} {Proceedings of the National Academy of Sciences}\
  }\textbf {\bibinfo {volume} {108}},\ \bibinfo {pages} {16549} (\bibinfo
  {year} {2011})}\BibitemShut {NoStop}%
\bibitem [{\citenamefont {Varma}\ \emph {et~al.}(1989)\citenamefont {Varma},
  \citenamefont {Littlewood}, \citenamefont {Schmitt-Rink}, \citenamefont
  {Abrahams},\ and\ \citenamefont {Ruckenstein}}]{VarmaMFL}%
  \BibitemOpen
  \bibfield  {author} {\bibinfo {author} {\bibfnamefont {C.~M.}\ \bibnamefont
  {Varma}}, \bibinfo {author} {\bibfnamefont {P.~B.}\ \bibnamefont
  {Littlewood}}, \bibinfo {author} {\bibfnamefont {S.}~\bibnamefont
  {Schmitt-Rink}}, \bibinfo {author} {\bibfnamefont {E.}~\bibnamefont
  {Abrahams}}, \ and\ \bibinfo {author} {\bibfnamefont {A.~E.}\ \bibnamefont
  {Ruckenstein}},\ }\href {\doibase 10.1103/PhysRevLett.63.1996} {\bibfield
  {journal} {\bibinfo  {journal} {Phys. Rev. Lett.}\ }\textbf {\bibinfo
  {volume} {63}},\ \bibinfo {pages} {1996} (\bibinfo {year}
  {1989})}\BibitemShut {NoStop}%
\bibitem [{\citenamefont {{Abanov}}\ \emph {et~al.}(2003)\citenamefont
  {{Abanov}}, \citenamefont {{Chubukov}},\ and\ \citenamefont
  {{Schmalian}}}]{chubukov_adv_phys}%
  \BibitemOpen
  \bibfield  {author} {\bibinfo {author} {\bibfnamefont {A.}~\bibnamefont
  {{Abanov}}}, \bibinfo {author} {\bibfnamefont {A.~V.}\ \bibnamefont
  {{Chubukov}}}, \ and\ \bibinfo {author} {\bibfnamefont {J.}~\bibnamefont
  {{Schmalian}}},\ }\href {\doibase 10.1080/0001873021000057123} {\bibfield
  {journal} {\bibinfo  {journal} {Advances in Physics}\ }\textbf {\bibinfo
  {volume} {52}},\ \bibinfo {pages} {119} (\bibinfo {year} {2003})}\BibitemShut
  {NoStop}%
\bibitem [{\citenamefont {Tsvelik}(2017)}]{tsvelik2017ladder}%
  \BibitemOpen
  \bibfield  {author} {\bibinfo {author} {\bibfnamefont {A.}~\bibnamefont
  {Tsvelik}},\ }\href@noop {} {\bibfield  {journal} {\bibinfo  {journal}
  {Physical Review B}\ }\textbf {\bibinfo {volume} {95}},\ \bibinfo {pages}
  {201112} (\bibinfo {year} {2017})}\BibitemShut {NoStop}%
\bibitem [{\citenamefont {Rice}\ \emph {et~al.}(2017)\citenamefont {Rice},
  \citenamefont {Robinson},\ and\ \citenamefont {Tsvelik}}]{rice2017umklapp}%
  \BibitemOpen
  \bibfield  {author} {\bibinfo {author} {\bibfnamefont {T.~M.}\ \bibnamefont
  {Rice}}, \bibinfo {author} {\bibfnamefont {N.~J.}\ \bibnamefont {Robinson}},
  \ and\ \bibinfo {author} {\bibfnamefont {A.~M.}\ \bibnamefont {Tsvelik}},\
  }\href@noop {} {\bibfield  {journal} {\bibinfo  {journal} {Physical Review
  B}\ }\textbf {\bibinfo {volume} {96}},\ \bibinfo {pages} {220502} (\bibinfo
  {year} {2017})}\BibitemShut {NoStop}%
\bibitem [{\citenamefont {Hartnoll}\ \emph {et~al.}(2018)\citenamefont
  {Hartnoll}, \citenamefont {Lucas},\ and\ \citenamefont
  {Sachdev}}]{hartnoll2018holographic}%
  \BibitemOpen
  \bibfield  {author} {\bibinfo {author} {\bibfnamefont {S.~A.}\ \bibnamefont
  {Hartnoll}}, \bibinfo {author} {\bibfnamefont {A.}~\bibnamefont {Lucas}}, \
  and\ \bibinfo {author} {\bibfnamefont {S.}~\bibnamefont {Sachdev}},\
  }\href@noop {} {\emph {\bibinfo {title} {Holographic quantum matter}}}\
  (\bibinfo  {publisher} {MIT press},\ \bibinfo {year} {2018})\BibitemShut
  {NoStop}%
\bibitem [{\citenamefont {Wu}\ \emph {et~al.}(2018)\citenamefont {Wu},
  \citenamefont {Chen}, \citenamefont {Jian}, \citenamefont {You},\ and\
  \citenamefont {Xu}}]{wu2018candidate}%
  \BibitemOpen
  \bibfield  {author} {\bibinfo {author} {\bibfnamefont {X.}~\bibnamefont
  {Wu}}, \bibinfo {author} {\bibfnamefont {X.}~\bibnamefont {Chen}}, \bibinfo
  {author} {\bibfnamefont {C.-M.}\ \bibnamefont {Jian}}, \bibinfo {author}
  {\bibfnamefont {Y.-Z.}\ \bibnamefont {You}}, \ and\ \bibinfo {author}
  {\bibfnamefont {C.}~\bibnamefont {Xu}},\ }\href@noop {} {\bibfield  {journal}
  {\bibinfo  {journal} {Physical Review B}\ }\textbf {\bibinfo {volume} {98}},\
  \bibinfo {pages} {165117} (\bibinfo {year} {2018})}\BibitemShut {NoStop}%
\bibitem [{\citenamefont {Patel}\ \emph {et~al.}(2018)\citenamefont {Patel},
  \citenamefont {McGreevy}, \citenamefont {Arovas},\ and\ \citenamefont
  {Sachdev}}]{patel2018magnetotransport}%
  \BibitemOpen
  \bibfield  {author} {\bibinfo {author} {\bibfnamefont {A.~A.}\ \bibnamefont
  {Patel}}, \bibinfo {author} {\bibfnamefont {J.}~\bibnamefont {McGreevy}},
  \bibinfo {author} {\bibfnamefont {D.~P.}\ \bibnamefont {Arovas}}, \ and\
  \bibinfo {author} {\bibfnamefont {S.}~\bibnamefont {Sachdev}},\ }\href@noop
  {} {\bibfield  {journal} {\bibinfo  {journal} {Physical Review X}\ }\textbf
  {\bibinfo {volume} {8}},\ \bibinfo {pages} {021049} (\bibinfo {year}
  {2018})}\BibitemShut {NoStop}%
\bibitem [{\citenamefont {Chowdhury}\ and\ \citenamefont
  {Berg}(2020)}]{ErezBerg20}%
  \BibitemOpen
  \bibfield  {author} {\bibinfo {author} {\bibfnamefont {D.}~\bibnamefont
  {Chowdhury}}\ and\ \bibinfo {author} {\bibfnamefont {E.}~\bibnamefont
  {Berg}},\ }\href {\doibase 10.1103/PhysRevResearch.2.013301} {\bibfield
  {journal} {\bibinfo  {journal} {Phys. Rev. Research}\ }\textbf {\bibinfo
  {volume} {2}},\ \bibinfo {pages} {013301} (\bibinfo {year}
  {2020})}\BibitemShut {NoStop}%
\bibitem [{\citenamefont {Paul}\ \emph {et~al.}(2013)\citenamefont {Paul},
  \citenamefont {P\'epin},\ and\ \citenamefont {Norman}}]{Paul13}%
  \BibitemOpen
  \bibfield  {author} {\bibinfo {author} {\bibfnamefont {I.}~\bibnamefont
  {Paul}}, \bibinfo {author} {\bibfnamefont {C.}~\bibnamefont {P\'epin}}, \
  and\ \bibinfo {author} {\bibfnamefont {M.~R.}\ \bibnamefont {Norman}},\
  }\href {\doibase 10.1103/PhysRevLett.110.066402} {\bibfield  {journal}
  {\bibinfo  {journal} {Phys. Rev. Lett.}\ }\textbf {\bibinfo {volume} {110}},\
  \bibinfo {pages} {066402} (\bibinfo {year} {2013})}\BibitemShut {NoStop}%
\bibitem [{\citenamefont {Merino}\ and\ \citenamefont
  {McKenzie}(2000)}]{McKenzieDMFT00}%
  \BibitemOpen
  \bibfield  {author} {\bibinfo {author} {\bibfnamefont {J.}~\bibnamefont
  {Merino}}\ and\ \bibinfo {author} {\bibfnamefont {R.~H.}\ \bibnamefont
  {McKenzie}},\ }\href {\doibase 10.1103/PhysRevB.61.7996} {\bibfield
  {journal} {\bibinfo  {journal} {Phys. Rev. B}\ }\textbf {\bibinfo {volume}
  {61}},\ \bibinfo {pages} {7996} (\bibinfo {year} {2000})}\BibitemShut
  {NoStop}%
\bibitem [{\citenamefont {Huang}\ \emph
  {et~al.}(2019{\natexlab{a}})\citenamefont {Huang}, \citenamefont {Sheppard},
  \citenamefont {Moritz},\ and\ \citenamefont {Devereaux}}]{huang19strange}%
  \BibitemOpen
  \bibfield  {author} {\bibinfo {author} {\bibfnamefont {E.~W.}\ \bibnamefont
  {Huang}}, \bibinfo {author} {\bibfnamefont {R.}~\bibnamefont {Sheppard}},
  \bibinfo {author} {\bibfnamefont {B.}~\bibnamefont {Moritz}}, \ and\ \bibinfo
  {author} {\bibfnamefont {T.~P.}\ \bibnamefont {Devereaux}},\ }\href@noop {}
  {\bibfield  {journal} {\bibinfo  {journal} {Science}\ }\textbf {\bibinfo
  {volume} {366}},\ \bibinfo {pages} {987} (\bibinfo {year}
  {2019}{\natexlab{a}})}\BibitemShut {NoStop}%
\bibitem [{\citenamefont {Michon}\ \emph {et~al.}()\citenamefont {Michon},
  \citenamefont {Berthod}, \citenamefont {Rischau}, \citenamefont {Ataei},
  \citenamefont {Chen}, \citenamefont {Komiya}, \citenamefont {Ono},
  \citenamefont {Taillefer}, \citenamefont {Marel},\ and\ \citenamefont
  {Georges}}]{Michon}%
  \BibitemOpen
  \bibfield  {author} {\bibinfo {author} {\bibfnamefont {B.}~\bibnamefont
  {Michon}}, \bibinfo {author} {\bibfnamefont {C.}~\bibnamefont {Berthod}},
  \bibinfo {author} {\bibfnamefont {C.~W.}\ \bibnamefont {Rischau}}, \bibinfo
  {author} {\bibfnamefont {A.}~\bibnamefont {Ataei}}, \bibinfo {author}
  {\bibfnamefont {L.}~\bibnamefont {Chen}}, \bibinfo {author} {\bibfnamefont
  {S.}~\bibnamefont {Komiya}}, \bibinfo {author} {\bibfnamefont
  {S.}~\bibnamefont {Ono}}, \bibinfo {author} {\bibfnamefont {L.}~\bibnamefont
  {Taillefer}}, \bibinfo {author} {\bibfnamefont {D.~V.~D.}\ \bibnamefont
  {Marel}}, \ and\ \bibinfo {author} {\bibfnamefont {A.}~\bibnamefont
  {Georges}},\ }\href@noop {} {\ }\BibitemShut {NoStop}%
\bibitem [{\citenamefont {Taheridehkordi}\ \emph {et~al.}(2019)\citenamefont
  {Taheridehkordi}, \citenamefont {Curnoe},\ and\ \citenamefont
  {LeBlanc}}]{AMI}%
  \BibitemOpen
  \bibfield  {author} {\bibinfo {author} {\bibfnamefont {A.}~\bibnamefont
  {Taheridehkordi}}, \bibinfo {author} {\bibfnamefont {S.~H.}\ \bibnamefont
  {Curnoe}}, \ and\ \bibinfo {author} {\bibfnamefont {J.~P.~F.}\ \bibnamefont
  {LeBlanc}},\ }\href {\doibase 10.1103/PhysRevB.99.035120} {\bibfield
  {journal} {\bibinfo  {journal} {Phys. Rev. B}\ }\textbf {\bibinfo {volume}
  {99}},\ \bibinfo {pages} {35120} (\bibinfo {year} {2019})}\BibitemShut
  {NoStop}%
\bibitem [{\citenamefont {Elazab}\ \emph {et~al.}(2022)\citenamefont {Elazab},
  \citenamefont {McNiven},\ and\ \citenamefont {LeBlanc}}]{libami}%
  \BibitemOpen
  \bibfield  {author} {\bibinfo {author} {\bibfnamefont {H.}~\bibnamefont
  {Elazab}}, \bibinfo {author} {\bibfnamefont {B.~D.~E.}\ \bibnamefont
  {McNiven}}, \ and\ \bibinfo {author} {\bibfnamefont {J.~P.~F.}\ \bibnamefont
  {LeBlanc}},\ }\href@noop {} {\bibfield  {journal} {\bibinfo  {journal}
  {arXiv}\ ,\ \bibinfo {pages} {2201.09868}} (\bibinfo {year}
  {2022})}\BibitemShut {NoStop}%
\bibitem [{\citenamefont {Jarrell}\ and\ \citenamefont
  {Gubernatis}(1996)}]{jarrell:maxent}%
  \BibitemOpen
  \bibfield  {author} {\bibinfo {author} {\bibfnamefont {M.}~\bibnamefont
  {Jarrell}}\ and\ \bibinfo {author} {\bibfnamefont {J.~E.}\ \bibnamefont
  {Gubernatis}},\ }\href {\doibase
  https://doi.org/10.1016/0370-1573(95)00074-7} {\bibfield  {journal} {\bibinfo
   {journal} {Physics Reports}\ }\textbf {\bibinfo {volume} {269}},\ \bibinfo
  {pages} {133} (\bibinfo {year} {1996})}\BibitemShut {NoStop}%
\bibitem [{\citenamefont {{Levy}}\ \emph {et~al.}(2017)\citenamefont {{Levy}},
  \citenamefont {{LeBlanc}},\ and\ \citenamefont {{Gull}}}]{maxent}%
  \BibitemOpen
  \bibfield  {author} {\bibinfo {author} {\bibfnamefont {R.}~\bibnamefont
  {{Levy}}}, \bibinfo {author} {\bibfnamefont {J.~P.~F.}\ \bibnamefont
  {{LeBlanc}}}, \ and\ \bibinfo {author} {\bibfnamefont {E.}~\bibnamefont
  {{Gull}}},\ }\href {\doibase 10.1016/j.cpc.2017.01.018} {\bibfield  {journal}
  {\bibinfo  {journal} {Comp. Phys. Comm.}\ }\textbf {\bibinfo {volume}
  {215}},\ \bibinfo {pages} {149} (\bibinfo {year} {2017})}\BibitemShut
  {NoStop}%
\bibitem [{\citenamefont {Sch\"ott}\ \emph {et~al.}(2016)\citenamefont
  {Sch\"ott}, \citenamefont {Locht}, \citenamefont {Lundin}, \citenamefont
  {Gr\aa{}n\"as}, \citenamefont {Eriksson},\ and\ \citenamefont
  {Di~Marco}}]{pade}%
  \BibitemOpen
  \bibfield  {author} {\bibinfo {author} {\bibfnamefont {J.}~\bibnamefont
  {Sch\"ott}}, \bibinfo {author} {\bibfnamefont {I.~L.~M.}\ \bibnamefont
  {Locht}}, \bibinfo {author} {\bibfnamefont {E.}~\bibnamefont {Lundin}},
  \bibinfo {author} {\bibfnamefont {O.}~\bibnamefont {Gr\aa{}n\"as}}, \bibinfo
  {author} {\bibfnamefont {O.}~\bibnamefont {Eriksson}}, \ and\ \bibinfo
  {author} {\bibfnamefont {I.}~\bibnamefont {Di~Marco}},\ }\href {\doibase
  10.1103/PhysRevB.93.075104} {\bibfield  {journal} {\bibinfo  {journal} {Phys.
  Rev. B}\ }\textbf {\bibinfo {volume} {93}},\ \bibinfo {pages} {075104}
  (\bibinfo {year} {2016})}\BibitemShut {NoStop}%
\bibitem [{\citenamefont {Fei}\ \emph {et~al.}(2021)\citenamefont {Fei},
  \citenamefont {Yeh},\ and\ \citenamefont {Gull}}]{gull:nevanlinna}%
  \BibitemOpen
  \bibfield  {author} {\bibinfo {author} {\bibfnamefont {J.}~\bibnamefont
  {Fei}}, \bibinfo {author} {\bibfnamefont {C.-N.}\ \bibnamefont {Yeh}}, \ and\
  \bibinfo {author} {\bibfnamefont {E.}~\bibnamefont {Gull}},\ }\href {\doibase
  10.1103/PhysRevLett.126.056402} {\bibfield  {journal} {\bibinfo  {journal}
  {Phys. Rev. Lett.}\ }\textbf {\bibinfo {volume} {126}},\ \bibinfo {pages}
  {56402} (\bibinfo {year} {2021})}\BibitemShut {NoStop}%
\bibitem [{\citenamefont {Nogaki}\ \emph {et~al.}(2023)\citenamefont {Nogaki},
  \citenamefont {Fei}, \citenamefont {Gull},\ and\ \citenamefont
  {Shinaoka}}]{Nogaki2023}%
  \BibitemOpen
  \bibfield  {author} {\bibinfo {author} {\bibfnamefont {K.}~\bibnamefont
  {Nogaki}}, \bibinfo {author} {\bibfnamefont {J.}~\bibnamefont {Fei}},
  \bibinfo {author} {\bibfnamefont {E.}~\bibnamefont {Gull}}, \ and\ \bibinfo
  {author} {\bibfnamefont {H.}~\bibnamefont {Shinaoka}},\ }\href {\doibase
  10.48550/ARXIV.2302.10476} {\enquote {\bibinfo {title} {Nevanlinna.jl: A
  julia implementation of nevanlinna analytic continuation},}\ } (\bibinfo
  {year} {2023})\BibitemShut {NoStop}%
\bibitem [{\citenamefont {Huang}\ \emph
  {et~al.}(2019{\natexlab{b}})\citenamefont {Huang}, \citenamefont {Sheppard},
  \citenamefont {Moritz},\ and\ \citenamefont {Devereaux}}]{Huang2019}%
  \BibitemOpen
  \bibfield  {author} {\bibinfo {author} {\bibfnamefont {E.~W.}\ \bibnamefont
  {Huang}}, \bibinfo {author} {\bibfnamefont {R.}~\bibnamefont {Sheppard}},
  \bibinfo {author} {\bibfnamefont {B.}~\bibnamefont {Moritz}}, \ and\ \bibinfo
  {author} {\bibfnamefont {T.~P.}\ \bibnamefont {Devereaux}},\ }\href {\doibase
  10.1126/science.aau7063} {\bibfield  {journal} {\bibinfo  {journal}
  {Science}\ }\textbf {\bibinfo {volume} {366}},\ \bibinfo {pages} {987}
  (\bibinfo {year} {2019}{\natexlab{b}})}\BibitemShut {NoStop}%
\bibitem [{\citenamefont {Vucicevic}\ \emph {et~al.}(2022)\citenamefont
  {Vucicevic}, \citenamefont {Predin},\ and\ \citenamefont
  {Ferrero}}]{Vucicevic2022}%
  \BibitemOpen
  \bibfield  {author} {\bibinfo {author} {\bibfnamefont {J.}~\bibnamefont
  {Vucicevic}}, \bibinfo {author} {\bibfnamefont {S.}~\bibnamefont {Predin}}, \
  and\ \bibinfo {author} {\bibfnamefont {M.}~\bibnamefont {Ferrero}},\ }\href
  {http://arxiv.org/abs/2208.04047} {\  (\bibinfo {year} {2022})}\BibitemShut
  {NoStop}%
\bibitem [{\citenamefont {Millis}\ and\ \citenamefont
  {Drew}(2003)}]{Millis2003}%
  \BibitemOpen
  \bibfield  {author} {\bibinfo {author} {\bibfnamefont {A.~J.}\ \bibnamefont
  {Millis}}\ and\ \bibinfo {author} {\bibfnamefont {H.~D.}\ \bibnamefont
  {Drew}},\ }\href {\doibase 10.1103/physrevb.67.214517} {\bibfield  {journal}
  {\bibinfo  {journal} {Physical Review B}\ }\textbf {\bibinfo {volume} {67}}
  (\bibinfo {year} {2003}),\ 10.1103/physrevb.67.214517}\BibitemShut {NoStop}%
\bibitem [{\citenamefont {Grissonnanche}\ \emph {et~al.}(2021)\citenamefont
  {Grissonnanche}, \citenamefont {Fang}, \citenamefont {Legros}, \citenamefont
  {Verret}, \citenamefont {Lalibert{\'{e}}}, \citenamefont {Collignon},
  \citenamefont {Zhou}, \citenamefont {Graf}, \citenamefont {Goddard},
  \citenamefont {Taillefer},\ and\ \citenamefont
  {Ramshaw}}]{Grissonnanche2021}%
  \BibitemOpen
  \bibfield  {author} {\bibinfo {author} {\bibfnamefont {G.}~\bibnamefont
  {Grissonnanche}}, \bibinfo {author} {\bibfnamefont {Y.}~\bibnamefont {Fang}},
  \bibinfo {author} {\bibfnamefont {A.}~\bibnamefont {Legros}}, \bibinfo
  {author} {\bibfnamefont {S.}~\bibnamefont {Verret}}, \bibinfo {author}
  {\bibfnamefont {F.}~\bibnamefont {Lalibert{\'{e}}}}, \bibinfo {author}
  {\bibfnamefont {C.}~\bibnamefont {Collignon}}, \bibinfo {author}
  {\bibfnamefont {J.}~\bibnamefont {Zhou}}, \bibinfo {author} {\bibfnamefont
  {D.}~\bibnamefont {Graf}}, \bibinfo {author} {\bibfnamefont {P.~A.}\
  \bibnamefont {Goddard}}, \bibinfo {author} {\bibfnamefont {L.}~\bibnamefont
  {Taillefer}}, \ and\ \bibinfo {author} {\bibfnamefont {B.~J.}\ \bibnamefont
  {Ramshaw}},\ }\href {\doibase 10.1038/s41586-021-03697-8} {\bibfield
  {journal} {\bibinfo  {journal} {Nature}\ }\textbf {\bibinfo {volume} {595}},\
  \bibinfo {pages} {667} (\bibinfo {year} {2021})}\BibitemShut {NoStop}%
\bibitem [{\citenamefont {Khurana}(1990)}]{Khurana1990}%
  \BibitemOpen
  \bibfield  {author} {\bibinfo {author} {\bibfnamefont {A.}~\bibnamefont
  {Khurana}},\ }\href {\doibase 10.1103/physrevlett.64.1990} {\bibfield
  {journal} {\bibinfo  {journal} {Physical Review Letters}\ }\textbf {\bibinfo
  {volume} {64}},\ \bibinfo {pages} {1990} (\bibinfo {year}
  {1990})}\BibitemShut {NoStop}%
\bibitem [{\citenamefont {Vucicevic}\ and\ \citenamefont
  {Zitko}(2021)}]{Vucicevic2021}%
  \BibitemOpen
  \bibfield  {author} {\bibinfo {author} {\bibfnamefont {J.}~\bibnamefont
  {Vucicevic}}\ and\ \bibinfo {author} {\bibfnamefont {R.}~\bibnamefont
  {Zitko}},\ }\href {\doibase 10.1103/PhysRevB.104.205101} {\bibfield
  {journal} {\bibinfo  {journal} {Physical Review B}\ }\textbf {\bibinfo
  {volume} {104}} (\bibinfo {year} {2021}),\
  10.1103/PhysRevB.104.205101}\BibitemShut {NoStop}%
\bibitem [{\citenamefont {Dumitrescu}\ \emph {et~al.}(2021)\citenamefont
  {Dumitrescu}, \citenamefont {Wentzell}, \citenamefont {Georges},\ and\
  \citenamefont {Parcollet}}]{Dumitrescu2021}%
  \BibitemOpen
  \bibfield  {author} {\bibinfo {author} {\bibfnamefont {P.~T.}\ \bibnamefont
  {Dumitrescu}}, \bibinfo {author} {\bibfnamefont {N.}~\bibnamefont
  {Wentzell}}, \bibinfo {author} {\bibfnamefont {A.}~\bibnamefont {Georges}}, \
  and\ \bibinfo {author} {\bibfnamefont {O.}~\bibnamefont {Parcollet}},\
  }\href@noop {} {\  (\bibinfo {year} {2021})}\BibitemShut {NoStop}%
\bibitem [{\citenamefont {Patel}\ \emph {et~al.}(2022)\citenamefont {Patel},
  \citenamefont {Guo}, \citenamefont {Esterlis},\ and\ \citenamefont
  {Sachdev}}]{Patel2022}%
  \BibitemOpen
  \bibfield  {author} {\bibinfo {author} {\bibfnamefont {A.~A.}\ \bibnamefont
  {Patel}}, \bibinfo {author} {\bibfnamefont {H.}~\bibnamefont {Guo}}, \bibinfo
  {author} {\bibfnamefont {I.}~\bibnamefont {Esterlis}}, \ and\ \bibinfo
  {author} {\bibfnamefont {S.}~\bibnamefont {Sachdev}},\ }\href
  {http://arxiv.org/abs/2203.04990} {\  (\bibinfo {year} {2022})}\BibitemShut
  {NoStop}%
\bibitem [{\citenamefont {Parcollet}\ \emph {et~al.}(1998)\citenamefont
  {Parcollet}, \citenamefont {Georges}, \citenamefont {Kotliar},\ and\
  \citenamefont {Sengupta}}]{Parcollet1998}%
  \BibitemOpen
  \bibfield  {author} {\bibinfo {author} {\bibfnamefont {O.}~\bibnamefont
  {Parcollet}}, \bibinfo {author} {\bibfnamefont {A.}~\bibnamefont {Georges}},
  \bibinfo {author} {\bibfnamefont {G.}~\bibnamefont {Kotliar}}, \ and\
  \bibinfo {author} {\bibfnamefont {A.}~\bibnamefont {Sengupta}},\ }\href@noop
  {} {\  (\bibinfo {year} {1998})}\BibitemShut {NoStop}%
\bibitem [{\citenamefont {Cha}\ \emph {et~al.}(2020)\citenamefont {Cha},
  \citenamefont {Patel}, \citenamefont {Gull},\ and\ \citenamefont
  {Kim}}]{Cha2020}%
  \BibitemOpen
  \bibfield  {author} {\bibinfo {author} {\bibfnamefont {P.}~\bibnamefont
  {Cha}}, \bibinfo {author} {\bibfnamefont {A.~A.}\ \bibnamefont {Patel}},
  \bibinfo {author} {\bibfnamefont {E.}~\bibnamefont {Gull}}, \ and\ \bibinfo
  {author} {\bibfnamefont {E.-A.}\ \bibnamefont {Kim}},\ }\href {\doibase
  10.1103/PhysRevResearch.2.033434} {\bibfield  {journal} {\bibinfo  {journal}
  {Physical Review Research}\ }\textbf {\bibinfo {volume} {2}} (\bibinfo {year}
  {2020}),\ 10.1103/PhysRevResearch.2.033434}\BibitemShut {NoStop}%
\bibitem [{\citenamefont {Kakehashi}\ and\ \citenamefont
  {Fulde}(2005)}]{Kakehashi2005}%
  \BibitemOpen
  \bibfield  {author} {\bibinfo {author} {\bibfnamefont {Y.}~\bibnamefont
  {Kakehashi}}\ and\ \bibinfo {author} {\bibfnamefont {P.}~\bibnamefont
  {Fulde}},\ }\href {\doibase 10.1103/PhysRevLett.94.156401} {\  (\bibinfo
  {year} {2005}),\ 10.1103/PhysRevLett.94.156401}\BibitemShut {NoStop}%
\bibitem [{\citenamefont {Virosztek}\ and\ \citenamefont
  {Ruvalds}(1990)}]{Virosztek1990}%
  \BibitemOpen
  \bibfield  {author} {\bibinfo {author} {\bibfnamefont {A.}~\bibnamefont
  {Virosztek}}\ and\ \bibinfo {author} {\bibfnamefont {J.}~\bibnamefont
  {Ruvalds}},\ }\href@noop {} {\bibfield  {journal} {\bibinfo  {journal}
  {PHYSICAL REVIEW B}\ }\textbf {\bibinfo {volume} {42}} (\bibinfo {year}
  {1990})}\BibitemShut {NoStop}%
\bibitem [{\citenamefont {Czycholl}(1991)}]{Czycholl1991}%
  \BibitemOpen
  \bibfield  {author} {\bibinfo {author} {\bibfnamefont {G.}~\bibnamefont
  {Czycholl}},\ }\href@noop {} {\bibfield  {journal} {\bibinfo  {journal} {Z.
  Phys. B-Condensed Matter}\ }\textbf {\bibinfo {volume} {83}},\ \bibinfo
  {pages} {93} (\bibinfo {year} {1991})}\BibitemShut {NoStop}%
\bibitem [{\citenamefont {W{\'{u}}}\ \emph {et~al.}(2022)\citenamefont
  {W{\'{u}}}, \citenamefont {Wang},\ and\ \citenamefont {Tremblay}}]{Wei2022}%
  \BibitemOpen
  \bibfield  {author} {\bibinfo {author} {\bibfnamefont {W.}~\bibnamefont
  {W{\'{u}}}}, \bibinfo {author} {\bibfnamefont {X.}~\bibnamefont {Wang}}, \
  and\ \bibinfo {author} {\bibfnamefont {A.-M.}\ \bibnamefont {Tremblay}},\
  }\href {\doibase 10.1073/pnas.2115819119} {\bibfield  {journal} {\bibinfo
  {journal} {Proceedings of the National Academy of Sciences}\ }\textbf
  {\bibinfo {volume} {119}} (\bibinfo {year} {2022}),\
  10.1073/pnas.2115819119}\BibitemShut {NoStop}%
\end{thebibliography}%

\pagebreak
\widetext
\begin{center}
\textbf{\large Supplementary material}
\end{center}
%%%%%%%%%% Merge with supplemental materials %%%%%%%%%%
%%%%%%%%%% Prefix a "S" to all equations, figures, and tables and reset the counter %%%%%%%%%%
\setcounter{equation}{0}
\setcounter{figure}{0}
\setcounter{table}{0}
\setcounter{page}{1}
\makeatletter
\renewcommand{\theequation}{S\arabic{equation}}
\renewcommand{\thefigure}{S\arabic{figure}}
\twocolumngrid
\section{Renormalized Perturbation expansion}\label{sec:appA}
We are interested in the half-filled Hubbard model on a 2D square lattice
\begin{equation}
H = \sum_{ ij \sigma} t_{ij}c_{i\sigma}^\dagger c_{j\sigma} + U\sum_{i} n_{i\uparrow} n_{i\downarrow},
\end{equation}
where $t_{ij}$ is the hopping amplitude, $c_{i\sigma}^{(\dagger)}$ ($c_{i\sigma}$) is the creation (annihilation) operator at site $i$, $\sigma \in \{\uparrow,\downarrow\}$ is the spin, $U$ is the onsite Hubbard interaction, $n_{i\sigma} = c_{i\sigma}^{\dagger}c_{i\sigma}$ is the number operator. The frequency-dependent longitudinal conductivity is related to the current-current correlation function
\begin{equation}
\sigma(\omega) = i\left(\chi_{jj}(0)-\chi_{jj}(\omega)\right)/\omega,\label{eq:S2}
\end{equation}
which is obtained in the linear response theory
\begin{equation}
\chi_{jj}(i\omega_n) = -\frac{1}{V}\int d\tau e^{i\omega_n\tau}\left\langle T_{\tau}j\left(\tau\right)j\left(0\right)\right\rangle,
\end{equation}
where $j = \sum_{k,\sigma} v_k \cdks{}{}\cks{}{}$ is the current operator and $v_k = \frac{\partial\epsilon_k}{\partial k}$ is the band velocity.
We compute $\chi_{jj}(i\omega_n)$ by performing a perturbative expansion with respect to the electron-electron interaction $U$. However, we do not start from the non-interacting $G_0\left(k,i\nu_n\right)$ but instead, use a renormalized perturbation scheme where the Green's function is dressed by including self-energies corrections
\begin{equation}
    G(k,i\nu_n)=\frac{1}{i\nu_n-\epsilon_{k}-\Sigma(k,i\nu_n)}.
\end{equation}
In particular, we use here the second-order self-energy correction to include a physical scattering time in the starting point of our expansion for the conductivity. The dressed Green's function is thus given by the Dyson equation
\begin{equation}
\includegraphics[width= 0.6 \linewidth]{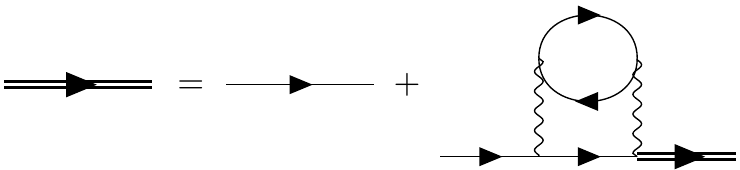},
\end{equation}
where the single line is the non-interacting $G_0\left(k,i\nu_n\right)$, the double line is the dressed $G(k,i\nu_n)$ and the wavy line represent the electron-electron interaction $U$. Note that there is no self-consistency in the calculation of the dressed Green's function in contrast to other ''bold diagrammatic`` approaches.\\

The expansion for the current-current correlation function is thus given by the skeleton series that excludes diagrams with self-energy insertions to avoid double counting
\begin{equation}
\includegraphics[width= \linewidth]{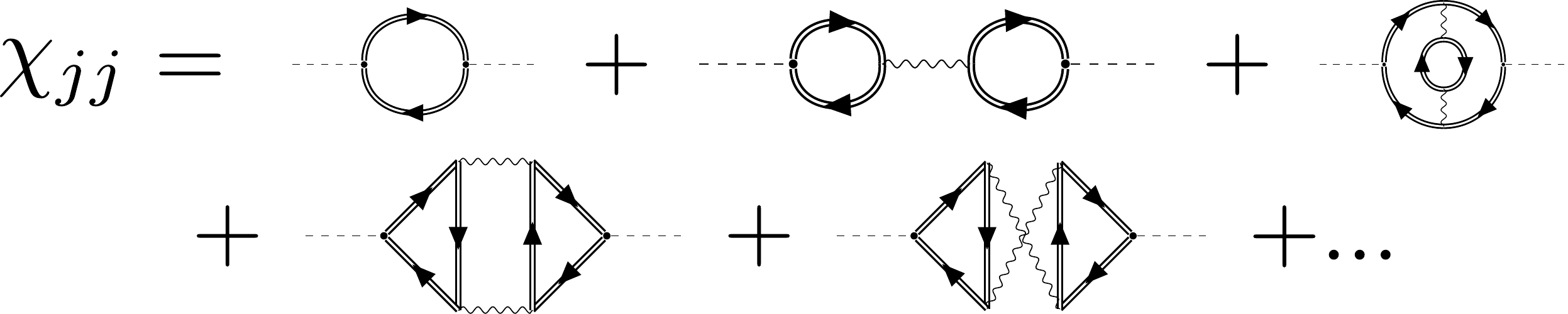} \label{eq:S6}
\end{equation}
This diagrammatic expansion is treated with Algorithmic Matsubara Integration (AMI) which allows for the symbolic summation over the internal Matsubara frequencies. As we don't have any analytic form for the self-energy that appears in the dressed Green's function, we use the spectral representation
\begin{equation}
    G(k,i\omega_n)=\int\limits_{-\infty}^{\infty}dx \frac{A(k,x)}{i\omega_n - x},
\end{equation}
where the spectral function $A(k,x)$ is directly related to the analytic continuation of the self-energy
\begin{equation}
    A(k,x)=\frac{-1}{\pi} \frac{Im\Sigma(k,x)}{(x-\epsilon_k-Re\Sigma(k,x))^2+(Im\Sigma(k,x))^2}.
\end{equation}
Here again, we use AMI to compute the self-energy on the real-frequency axis without the need for numerical analytical continuation on a grid in momentum space. In practice, the self-energy is computed using an analytic continuation parameter $i\nu_n = \nu+i\Gamma$ with $\Gamma = 1e-2$. \\
\begin{figure}
    \centering
    \includegraphics[width=\linewidth]{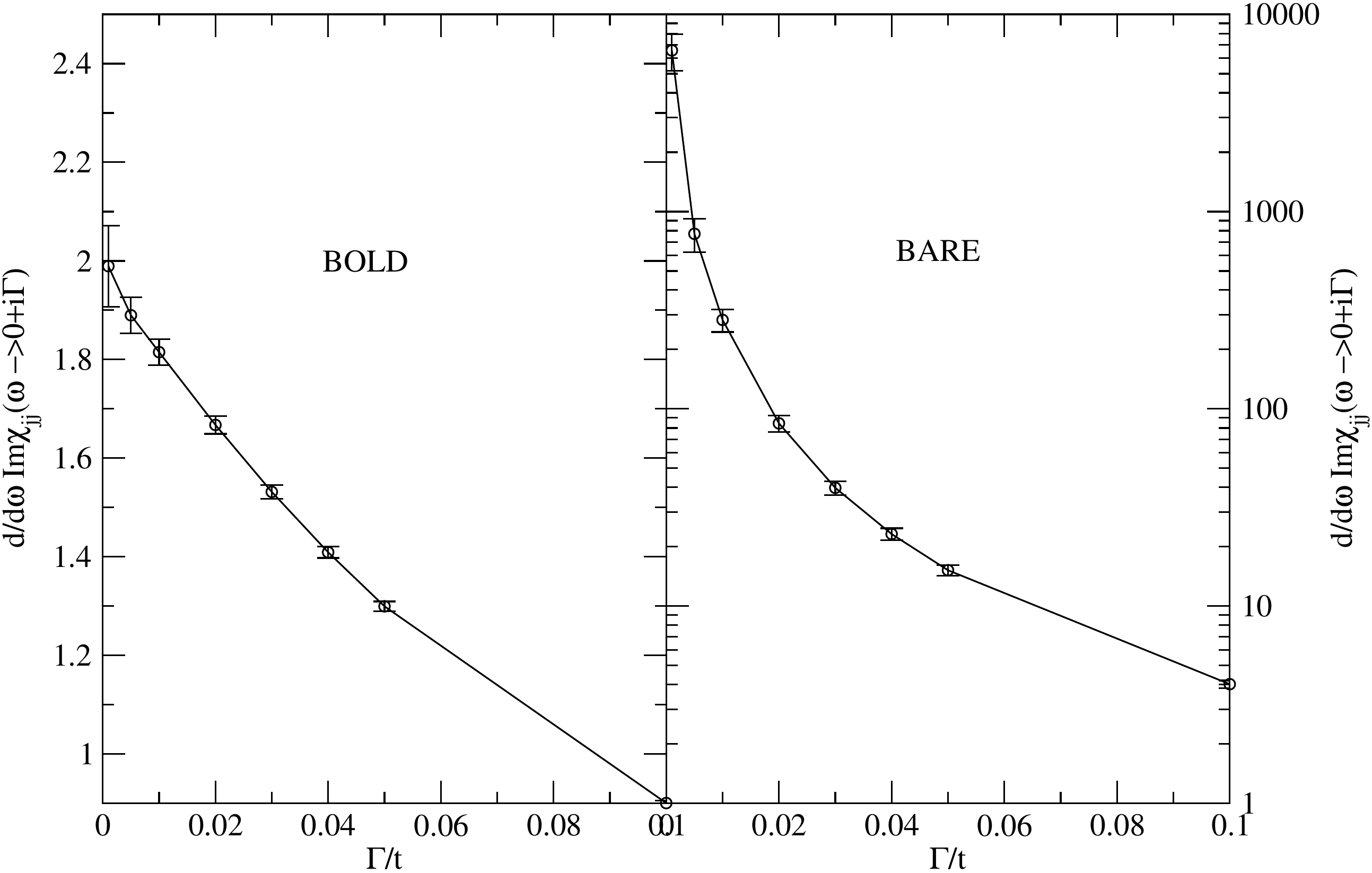}
    \caption{ \label{fig:S1} Comparison of the DC conductivity when changing the analytic continuation parameter $\Gamma$ between the renormalized (left) and the bare (right) expansion for the current-current correlation function. We used $U=$, $\beta=$}
\end{figure}
An example of the type of expressions generated by AMI can be obtained by looking at the lowest order diagram in Eq.\eqref{eq:S6}
\begin{align}
\chi_{jj}^{(0)}\left(i\omega_n\right) &= \frac{-2}{\beta N}\sum_{k,i\nu_n} G(k,i\nu_n)G(k,i\nu_n+i\omega_n) \nonumber \\
&= \frac{-2}{\beta N}\sum_{k,i\nu_n}\int\limits_{-\infty}^{\infty}dxdy \frac{A(k,x)A(k,y)}{\left(i\nu_n - x\right)\left(i\nu_n+i\omega_n - y\right)} \nonumber \\
&= \frac{-2}{N}\sum_{k}\int\limits_{-\infty}^{\infty}dxdy \frac{A(k,x)A(k,y)\left(n_f\left(x\right)-n_f\left(y\right)\right)}{i\omega_n +x-y},
\end{align}
where the remaining integrations are done using standard Monte-Carlo integration schemes.

This renormalized perturbation scheme has the advantage of drastically reducing the dependence of the result on the analytic continuation parameter used as the electronic self-energy acts as a regulator itself. An example is given in Fig.\ref{fig:S1} where we compare the imaginary part of the current-current correlation obtained using the renormalized expansion (left) to the one obtained with the standard bare expansion (right) in the dc ($\omega \to 0$) limit as we change the value of the analytic continuation parameter $\Gamma$. While the bare expansion diverges as $\Gamma \to 0$, the renormalized expansion converges to a finite value which is the dc conductivity.

Note that we can also get rid of the analytic continuation parameter entirely as the additional integrations can be used to resolve the $\delta$-function arising when taking the limit $i\omega_n = \omega+i0^+$
\begin{align}
\text{Im}\left[\chi_{jj}^{(0)} \left(\omega+i0^+\right)\right]=  \frac{2\pi}{N}\sum_{k} \int\limits_{-\infty}^{\infty}&dx A(k,x)A(k,x+\omega) \nonumber \\
& \times \left(n_f\left(x\right)-n_f\left(x+\omega\right)\right).
\end{align}
This is the method we use for all frequency-dependent results at leading order while we keep an analytic continuation parameter $\Gamma=1e-2$ in the calculation of the second-order results (vertex corrections) for numerical purposes.

\begin{figure}
    \centering
    \includegraphics[width=0.9 \linewidth]{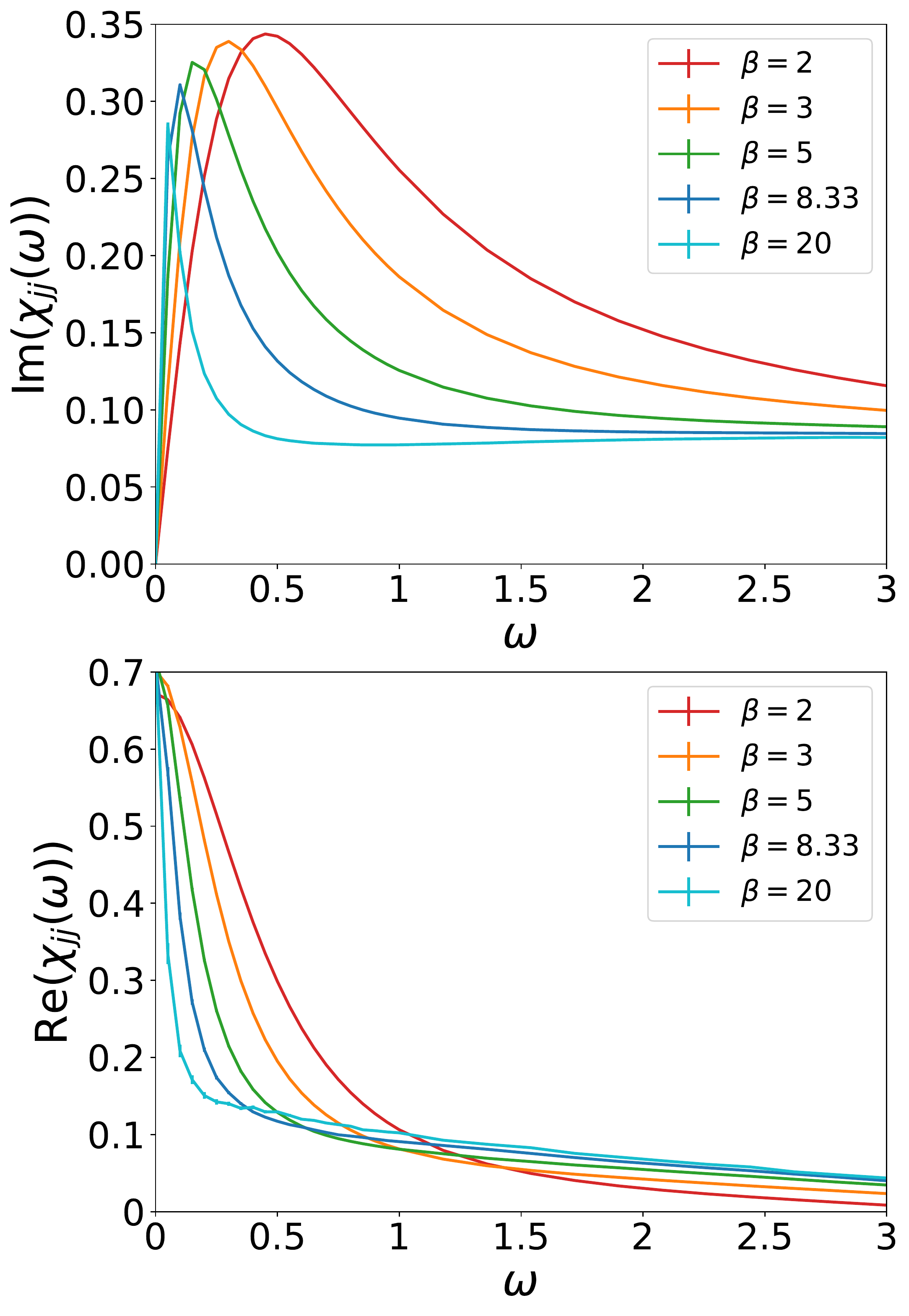}
    \caption{ \label{fig:S2} Frequency dependence of the imaginary (top) and real (bottom) part of the current-current correlation function at leading order. We used $U=3$ and $\Gamma \to 0$ (see text).}
\end{figure}

\section{Temperature dependence of the current-current correlation function}\label{sec:appB}

We present here the direct result of the calculation of the current-current calculation for the different temperatures $\beta = 2,3,5,8.33$ and $20$. The frequency dependence of the real and imaginary part of the current-current correlation function at leading ($0^{\text{th}}$ order in interaction) are shown in Fig.\ref{fig:S2}, error bars are smaller than the line width if not shown. As mentioned previously, the analytic continuation parameter $\Gamma=0$ is used. The calculation has been done up to $\omega=10$ but is not shown here for clarity.\\

\begin{figure}
    \centering
    \includegraphics[width=0.9 \linewidth]{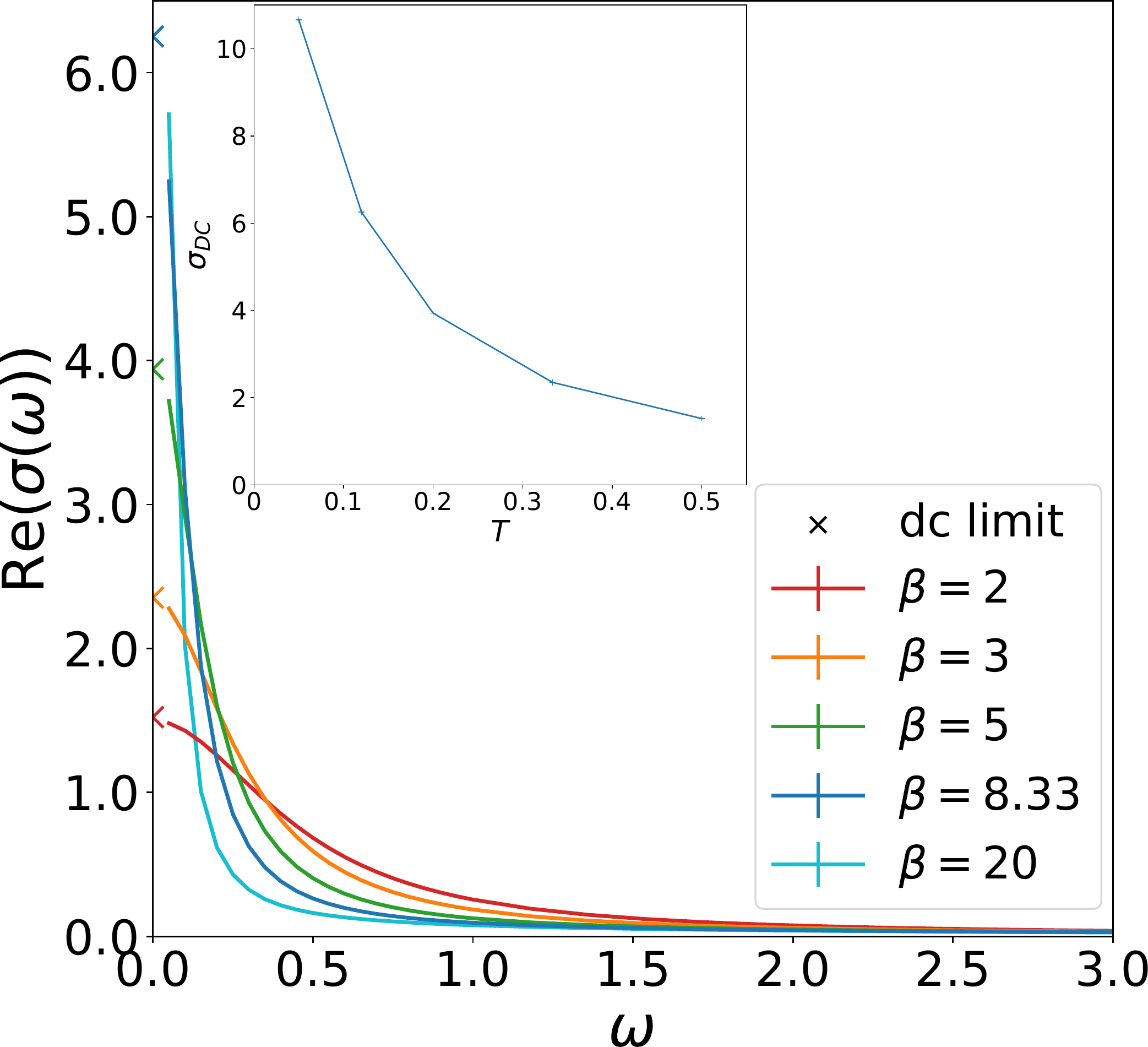}
    \caption{ \label{fig:S3} Real part of the conductivity obtained from $\chi_{jj}$ shown in Fig.\ref{fig:S2} for $\omega \neq 0$. The crosses indicate the value for $\sigma_{DC}=\sigma\left(\omega \to 0\right)$ of the corresponding colours are obtained using Eq.\eqref{eq:S11}. The inset shows the temperature dependence of $\sigma_{DC}$.}
\end{figure}

The frequency-dependent longitudinal conductivity can then directly be obtained using Eq.\eqref{eq:S2} for $\omega \neq 0$. The results for the real part of the conductivity are shown in Fig.\ref{fig:S3}. The conductivity has a Drude-like form with a temperature-dependent width and height. These are the results presented in Fig.1 of the main text although using the modulus and phase of the conductivity instead. 

The dc conductivity $\sigma\left(\omega \to 0\right)$ is obtained by taking symbolically the derivative of the current-current correlation function before the integration and evaluating it at $\omega = 0$,
\begin{equation}
\sigma\left(\omega \to 0\right) = -i\left. \frac{d\chi_{jj}\left(\omega\right)}{d\omega}\right|_{\omega=0}.\label{eq:S11}
\end{equation}
The results are shown by the corresponding crosses at $\omega=0$ and the inset shows the $1/T$ temperature dependence of the dc conductivity which is equivalent to the linear-in-$T$ resistivity reported in other numerical studies in the Hubbard model. This temperature dependence is also recovered in the scaling of the scattering time at $\omega=0$ shown in the main text (Fig.1).

\section{Scaling of the conductivity with $U=4$}\label{sec:appC}
\begin{figure}
	\centering
	\includegraphics[width=0.9 \linewidth]{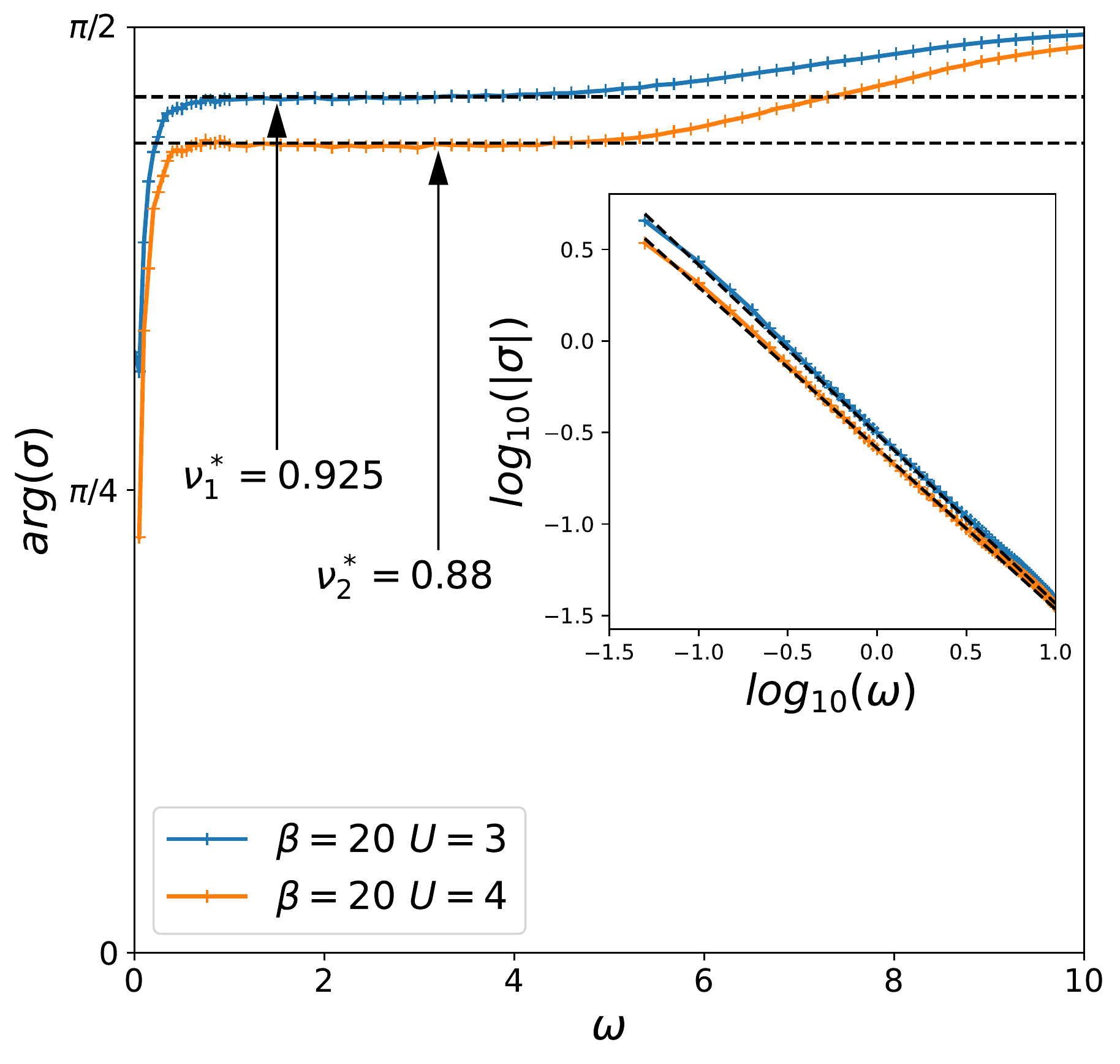}
	\caption{Frequency dependence of the phase of the conductivity at a fixed temperature $\beta=20$ obtained for two different values of the Hubbard interaction $U$. The value of the effective exponent $\nu^*$ decreases with the interaction strength according to Eq.\eqref{eq:nustar}. The inset shows the corresponding modulus of the conductivity that has a power law behaviour with an exponent $\nu^*$ similar to the one extracted from the phase plateau.}
	\label{fig:S4}
\end{figure}

Studies of the Planckian model have shown that the power law exponent observed in the modulus of the conductivity can be obtained analytically\cite{Michon}. Using the Planckian model (Eq.4 of the main text) with $\nu=1$, 
\begin{align}
&\text{Im}\left[ \Sigma \left( \omega \right) \right] = -\frac{g\pi}{\beta} \ f \left( \beta \omega \right),
\end{align}
the exponent $\nu^*$ can be shown\cite{Michon} to only depends on the coupling constant $g$,
\begin{equation}
\nu^* = 1-\frac{2g\left[1+2g\left(1+\ln 4\right)\right]}{\pi^2 g^2 + \left[1+2g\left(1+\ln 4\right)\right]^2}. \label{eq:nustar}
\end{equation}
This relation between the frequency dependence of the conductivity and the coupling constant can be verified in our study of the Hubbard model as shown in Fig.\ref{fig:S4}. 

We can extract the exponent $\nu^*$ by identifying the plateau reached by the phase of the conductivity for intermediate frequencies. Changing the value of the Hubbard interaction that enters in the expression of the second-order self-energy $\Sigma = U^2 \Sigma^{(2)}$ leads to a reduction of the effective exponent. We can extract the effective coupling constant from the data obtained at $U=3$, giving $g\left(U=3\right) \sim 0.047$. If we take the coupling constant $g$ to be proportional to the Hubbard interaction we then expect $g\left(U=4\right) = 0.047 \times \left(\frac{4}{3}\right)^2$ when taking the Hubbard interaction to be $U=4$. If the relation Eq.\eqref{eq:nustar} is respected we should then find $\nu^*\left(U=4\right) \sim 0.88$ which is what we observe in Fig.\ref{fig:S4}.

\begin{figure}
	\centering
	\includegraphics[width=0.9 \linewidth]{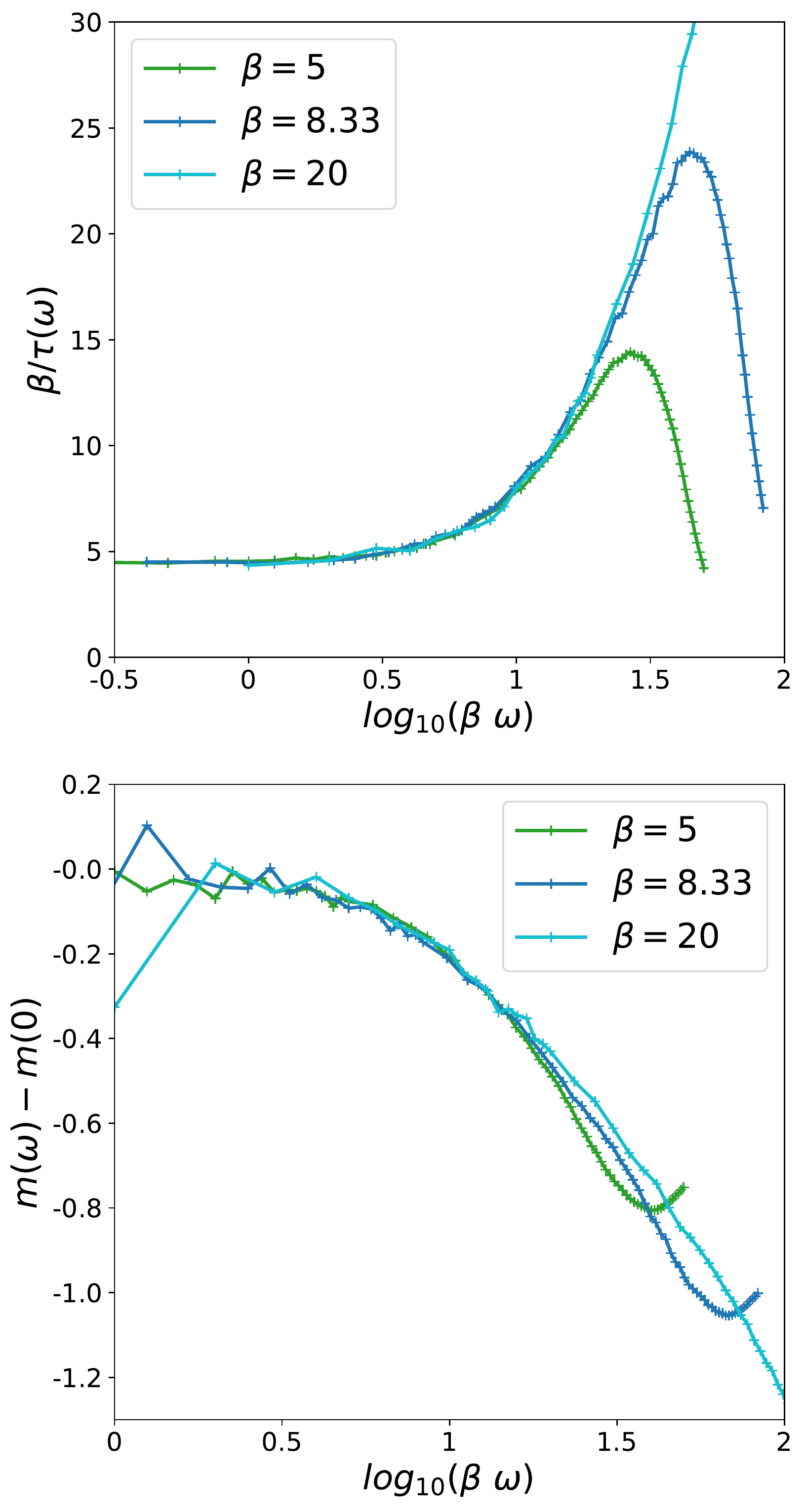}
	\caption{$\bm{(c)}$ inverse scattering time and $\bm{(d)}$ effective mass extracted from the conductivity when rewritten as an extended Drude model given in Eq.\eqref{eq:S14} for $U=4$.}
	\label{fig:S5}
\end{figure}

\newpage
We can also check that the scaling properties described in the main text remain when we change the value of the Hubbard interaction. The scattering time $\tau\left(\omega\right)$ and effective mass $m\left(\omega\right)/m_0$  extracted form the conductivity with $U=4$ when using a generalized Drude model,
\begin{equation}
\frac{1}{\tau\left(\omega\right)} = \text{Re}\left[\frac{1}{\sigma\left(\omega\right)}\right], \qquad
\frac{m^*\left(\omega\right)}{m_0} = -\text{Im}\left[\frac{1}{\omega \sigma\left(\omega\right)}\right] \label{eq:S14},
\end{equation}
are shown in Fig.\ref{fig:S5} for the lowest temperatures. We can see that we recover the temperature scaling observed at $U=3$ which we described in the main text.
\\

Note that because we restricted our calculation to the second-order self-energy and to the leading order for the current-current calculation we do not expect it to hold in the moderate to strong coupling regime. The results presented here should however be exact in the weak coupling regime when the higher order terms we neglected are suppressed and the fact that the scaling properties discussed are independent of the interaction strength means that we expect the Planckian behaviour to occur in the weakly coupled Hubbard model.

\end{document}